\documentclass[technote,9pt]{IEEEtran}
\usepackage{amsthm}
\usepackage{url}
\usepackage[cp1255]{inputenc}
\usepackage{amsfonts}
\usepackage{amssymb}
\usepackage{amsmath}
\usepackage{verbatim}
\usepackage{graphics}
\usepackage{graphicx}
\usepackage{bbm}
\usepackage{cancel}
\usepackage{fancyhdr}
\usepackage{cite}

\let\ORIGRightarrow=\Rightarrow

\let\Rightarrow=\ORIGRightarrow

\usepackage[usenames]{color}

\usepackage{algorithmic}
\usepackage{algorithm}
\newcommand{\E}{\mathbb{E}} 
\newcommand{\EE}[1]{\mathbb{E}\left[#1\right]}

\DeclareMathOperator{\diag}{diag}

\newcommand{\gaus}[2]{{\mathcal{N}}\paren{#1,#2}}

\newcommand{\reals}{\mathbb{R}}

\newcommand{\cM}{\mathcal{M}}

\newcommand{\beq}{\begin{equation}}
\newcommand{\eeq}{\end{equation}}

\newcommand{\beqa}{\begin{eqnarray}}
\newcommand{\eeqa}{\end{eqnarray}}
\newcommand{\non}{\nonumber}

\newcommand{\bm}[1]{\mbox{\boldmath{$#1$}}}

\newcommand{\set}[1]{\left\{#1\right\}}

\newcommand{\paren}[1]{\left(#1\right)}

\newcommand{\T}{\top}

\newcommand{\bx}{{{x}}}
\newcommand{\by}{{{y}}}

\newcommand{\bv}{{{v}}}
\newcommand{\bw}{{{w}}}

\usepackage[font=bf]{subfig}
\usepackage{mathdots}

\DeclareGraphicsExtensions{.eps}
\graphicspath{{../figures/}}

\newcommand{\Gnom}{G_{\rm{nom}}}

\newcommand{\Hnom}{H_{\rm{nom}}}

\newcommand{\Gcl}{G_{\rm{cl}}}

\newcommand{\Si}{\Sigma}
\newcommand{\La}{\Lambda}
\newcommand{\Up}{\Upsilon}

\newtheorem{coro}{Corollary}

\newtheorem{lemm}{Lemma}
\newtheorem*{lemm*}{Lemma}

\begin{document}
%

\title{
LMMSE Filtering in Feedback Systems with White Random Modes: Application to Tracking in Clutter
}

\author{Daniel~Sigalov, Tomer~Michaeli,
        and Yaakov~Oshman,~\IEEEmembership{Fellow,~IEEE}
}

\markboth{Submitted to IEEE Trans. on Automatic Control}%
{Shell \MakeLowercase{\textit{et al.}}: Bare Demo of IEEEtran.cls for Journals}

\maketitle

\begin{abstract}
A generalized state space representation of dynamical systems with random modes switching according to a white random process is presented. The new formulation includes a term, in the dynamics equation, that depends on the most recent linear minimum mean squared error (LMMSE) estimate of the state. This can model the behavior of a feedback control system featuring a state estimator. The measurement equation is allowed to depend on the previous LMMSE estimate of the state, which can represent the fact that measurements are obtained from a validation window centered about the predicted measurement and not from the entire surveillance region. 
The LMMSE filter is derived for the considered problem. The approach is demonstrated in the context of target tracking in clutter and is shown to 
be competitive with several popular nonlinear methods.
\end{abstract}


\section{Introduction}\label{section:intro}
State estimation in dynamical systems with randomly switching
coefficients is an important problem in many applications. Natural examples are maneuvering target tracking and fault
detection and isolation algorithms, featured, e.g., in aerospace
navigation systems. In the standard modeling the dynamics
of the continuously-valued state, and, possibly, its measurement
equation, are controlled by a discrete evolving mode. This is the well known concept of hybrid
systems~\cite{boukas:liu:02}.


Various problems have been formulated using the hybrid systems
framework. In problems involving uncertain observations, such
as~\cite{Nahi,Hadidi}, the mode affects the matrices of the measurement equation.
In target tracking applications, considered in,
e.g.,~\cite{nahi_parameters,ackerson1970ses,blom_YBS_IMM},
the mode usually affects the dynamics equation.

We consider a state space representation of dynamical systems with
random coefficients that constitute a white stochastic sequence, accompanied by the following feedback terms.
First, we allow the system input to depend on the latest estimate of the state, as is common
practice in closed loop control systems. In this work, the state
estimate is taken to be the linear minimum mean squared error (LMMSE)
estimate.
In addition, the measurement equation is also set to depend on the latest LMMSE
state estimate. This can represent the fact that observations are not taken in the
entire feasible space, but, rather, in a small validation window set
about the predicted measurement of the state.

It is well known~\cite{ackerson1970ses} that, even for the case of independently switching modes, the optimal estimate of the state cannot be obtained without resorting to exhaustive
enumeration. Therefore, significant efforts have been dedicated to
developing suboptimal approaches for state estimation in hybrid
systems and especially for the important subclass of jump linear
systems (JLS). The most popular nonlinear methods include
the generalized pseudo-Bayesian (GPB) filter~\cite{ackerson1970ses} and the
interacting multiple model (IMM) algorithm~\cite{blom_YBS_IMM}.
Alternatively, one may consider optimality within the narrower family of linear
filters. Among these we mention~\cite{Nahi} and~\cite{Hadidi} that considered estimation with
uncertain observations,~\cite{dekoning:stochastic:1984} that derived a Kalman filter-like (KF)
algorithm for a JLS with independently switching modes and uncorrelated matrices within each time step, and~\cite{costa1994lmm} that derived an LMMSE scheme for a Markov JLS by means of state augmentation.
In addition, in some cases, parts of the state may be estimated optimally while others in a linear optimal manner, 
as was shown in~\cite{michaeli:PLMMSE:2012}.

In this paper we concentrate on feedback JLS with independent mode
transitions and consider optimal estimation within the family of linear filters.
We derive a recursive LMMSE algorithm that may be conveniently implemented
in a recursive form, eliminating the need for unbounded memory. Unlike~\cite{dekoning:stochastic:1984}, we do not assume that the matrices within each time step are uncorrelated. This allows tackling a wider variety of problems, such as tracking in clutter, which cannot be modeled directly within the framework of~\cite{dekoning:stochastic:1984}. On the other hand, since we still treat the easier case of independent, rather than Markov, mode transitions, we do not require
state augmentation, as does the algorithm of~\cite{costa1994lmm}. Our filter reduces to several previously reported results when the parameters of the underlying problem are appropriately adjusted.
As an illustration, we formulate the problem of target tracking in clutter within the proposed framework and show
that the resulting filter is competitive with several classical nonlinear methods.

The paper is organized as follows. In Sec.~\ref{section:model} we describe the proposed modeling and survey some related work. The recursive LMMSE algorithm is derived in Sec.~\ref{section:derivation}. An application to target tracking in clutter, followed by a numerical study, is presented in Sec.~\ref{section:example1}.
Concluding remarks are given in Sec.~\ref{section:conclusion}.

\section{System Model and Related Work}\label{section:model}
We consider the dynamical system
\begin{subequations}\label{Eq:problem:input:system}
  \begin{align}
    \label{Eq:problem:input:dynamics}
    \bx_{k+1} & = A_k\bx_k + B_ku_k  + C_k\bw_k \\    \label{Eq:problem:input:measurement}
    \by_k & = H_k\bx_k + G_k\bv_k + F_k\hat{\bx}_{k-1} ,
\end{align}
\end{subequations}
where $\bx_k\in\reals^n$ and $\by_k\in\reals^m$ are the state and
measurement vectors at time $k$, respectively. The processes
$\set{\bw_k}$ and $\set{\bv_k}$ constitute zero-mean unity-covariance
strictly white sequences, and $\bx_0$ is a random vector (RV) with mean
$\bar{\bx}_0$ and second-order moment $P_0$.

We consider two variants for the modeling of  $u_k$.
In the first case, $u_k$ is a known deterministic
input. However, because in some cases $u_k$ serves as a closed loop control
signal, it is common practice to let it depend on the most recent
estimate of the state. Thus, in the second variant we set
$u_k=\hat{x}_k$, where $\hat{x}_k$ is the LMMSE estimate of $x_k$
using the measurement history $\mathcal{Y}_{k}\triangleq\set{\by_1,\ldots,\by_{k}}$.

Likewise, the term $\hat{\bx}_{k-1}$ in the measurement equation is
the LMMSE estimate of $\bx_{k-1}$ based on the measurement history
$\mathcal{Y}_{k-1}$. Affecting the measurement at time $k$, the term
$F_k\hat{\bx}_{k-1}$ can be used to represent the fact that
observations are not taken in the entire space, but, rather, in a small validation window, set about the
predicted measurement.

The system mode, $\cM_k \triangleq\set{A_k,B_k,C_k,H_k,G_k,F_k}$,
is a strictly white random process with known distribution. The quantities $\set{\bw_k}$,
$\set{\bv_k}$, $\set{\cM_k}$, and $\bx_0$ are assumed to be independent.

We seek to obtain the LMMSE estimate $\hat{\bx}_{k+1}$ using
the measurements $\mathcal{Y}_{k+1}$. It will be shown in the sequel
that, in our setting,  $\hat{\bx}_{k+1}$ conveniently possesses the
recursive form
\begin{align}\label{Eq:problem:recursive}
\hat{\bx}_{k+1}=L_k\hat{\bx}_{k}+K_k\by_{k+1}+J_ku_k
\end{align}
%
thus avoiding the need to store the entire measurement sequence. When $u_k=\hat{x}_k$,
the terms $L_k\hat{\bx}_{k}$ and $J_k\hat{\bx}_{k}$
in~\eqref{Eq:problem:recursive} may be grouped together.

Note that the described problem does not require the system mode to
assume values in a discrete domain as opposed to, e.g.~\cite{Nahi,Hadidi,costa1994lmm}.
In addition, the above formulation allows evolution not only of the entries of the mode matrices, but also of their dimensions~\cite{yuan2012mimm}. This observation allows treatment of problems that, to the best of our knowledge, have not been previously considered in the context of LMMSE algorithms. One such example is given in Section~\ref{section:example1}.


For the setting without feedback terms, 
several variants and special cases of the
presented problem have been considered in the past. Independent measurement faults were treated, in an LMMSE sense, in~\cite{Nahi}.
De Koning~\cite{dekoning:stochastic:1984} considered a more general
case of independently switching modes where, however, the mode elements are assumed uncorrelated, and Costa~\cite{costa1994lmm} developed, by means of state augmentation, a recursive LMMSE filter for systems with discrete modes obeying Markov dynamics.
Additional contributions include~\cite{Hadidi}, that considered correlated faults,~\cite{jackson1976ole}, that allowed correlations between subsequent fault variables,
and~\cite{nahi_parameters}, that proposed an LMMSE filter for the static multiple model problem~\cite{magill1965oae}.
Related nonlinear solutions were proposed in~\cite{ackerson1970ses,blom_YBS_IMM,sigalov:oshman:fusion10}
and references therein.

Besides the novel introduction of the feedback terms, this paper contains several additional contributions. First, we derive a recursive LMMSE algorithm without assuming uncorrelatedness of the mode elements, as done in~\cite{dekoning:stochastic:1984}. This assumption precludes the utilization of the algorithm of~\cite{dekoning:stochastic:1984} even for the simple problem of uncertain observations where measurement noise has a higher variance when faults occur, not to mention more involved settings, such as tracking in clutter. In addition, our algorithm is derived without state augmentation and without assuming discrete modes, as done in~\cite{costa1994lmm}. Finally, the approach allows a broader class of problem to be formulated within a single state-space model. Specifically, the new feedback terms allow the application of the idea to the problem of tracking in clutter.



\section{Linear Optimal Recursive Estimation}\label{section:derivation}
\noindent
We begin the derivation with deterministic $u_k$. The stochastic case is treated in Section~\ref{section:random:inputs}.

Let $Y_k$ be the RV obtained by concatenating the elements of $\mathcal{Y}_k$. We derive the result using the following lemma, which follows from~\cite[p.~190]{mendel1986lessons} and the linearity of the MMSE estimator in the Gaussian case.
%
\begin{lemm*}\label{Lemma:derivation2:recursive}
Let $x$, $y$ and $z$ be RVs and let $\hat{x}(z)$ and $\hat{x}(y,z)$ denote, respectively, the LMMSE estimates of $x$ using $z$, and using both $y$ and $z$. Let $\hat{y}(z)$ be the LMMSE estimate of $y$ using $z$. Then
$\hat{x}(y,z)
    =\hat{x}(z)+\Gamma_{x\tilde{y}}\Gamma_{\tilde{y}\tilde{y}}^{-1}\tilde{y},
$ 
where $\tilde{y}=y-\hat{y}(z)$ and $\Gamma_{ab}$ is the cross-covariance matrix between the RVs $a$ and $b$.
\end{lemm*}
\noindent
Letting $z\triangleq Y_{k}$, $y\triangleq y_{k+1}$ and using the lemma, 
the LMMSE estimate of ${x}_{k+1}$ using $\mathcal{Y}_{k+1}$ is
\begin{align}\label{Eq:derivation2:recursive:thru:lemma}
\hat{x}_{k+1}
    &=\hat{x}_{k+1}^{-}+\Gamma_{x_{k+1}\tilde{y}_{k+1}}\Gamma_{\tilde{y}_{k+1}\tilde{y}_{k+1}}^{-1}\tilde{y}_{k+1},
\end{align}
where $\hat{x}_{k+1}^{-}$ is the LMMSE estimate of $x_{k+1}$ using $\mathcal{Y}_k$, $\tilde{y}_{k+1}\triangleq y_{k+1}-\hat{y}_{k+1}^-$, and
$\hat{y}_{k+1}^{-}$ is the LMMSE estimate of $y_{k+1}$ using
$\mathcal{Y}_k$. If $\Gamma_{\tilde{y}_{k+1}\tilde{y}_{k+1}}$ is singular the lemma still holds with the inverse replaced by the Moore-Penrose pseudo-inverse. It is easily verified that
\begin{align}\label{Eq:derivation2:time:update}
\hat{x}_{k+1}^{-}
    &=\EE{A_k}\hat{x}_k+\EE{B_k}u_k\\\non
\hat{y}_{k+1}^{-}
    &=\EE{H_{k+1}}\hat{x}_{k+1}^{-}+\EE{F_{k+1}}\hat{x}_k\\\label{Eq:derivation2:time:update:y}
    &=(\EE{H_{k+1}}\EE{A_k}\!+\!\EE{F_{k+1}})\hat{x}_k\!+\!\EE{H_{k+1}}\EE{B_k}u_k.
\end{align}
Plugging~\eqref{Eq:derivation2:time:update} in~\eqref{Eq:derivation2:recursive:thru:lemma} we identify the desired matrix coefficients $K_k$, $L_k$, and $J_k$ of~\eqref{Eq:problem:recursive} as follows:
\begin{align}\label{Eq:derivation2:recursive:Kk}
K_k&=\Gamma_{x_{k+1}\tilde{y}_{k+1}}\Gamma_{\tilde{y}_{k+1}\tilde{y}_{k+1}}^{-1}\\\label{Eq:derivation2:recursive:Lk}
L_k&=(I-K_k\EE{H_{k+1}})\EE{A_k}-K_k\EE{F_{k+1}}\\\label{Eq:derivation2:recursive:Jk}
J_k&=(I-K_k\EE{H_{k+1}})\EE{B_k}.
\end{align}
We now compute the covariance terms $\Gamma_{x_{k+1}\tilde{y}_{k+1}}$ and $\Gamma_{\tilde{y}_{k+1}\tilde{y}_{k+1}}$.
\subsection{Computation of $\Gamma_{x_{k+1}\tilde{y}_{k+1}}$}
\noindent
Since $\hat{y}_{k+1}^{-}$ is unbiased, and using~\eqref{Eq:problem:input:measurement} and~\eqref{Eq:derivation2:time:update:y},
\begin{align}\non
\Gamma_{x_{k+1}\tilde{y}_{k+1}}
    &=\EE{x_{k+1}(y_{k+1}-\hat{y}_{k+1}^-)^\T}\\\non
    &=\EE{\bx_{k+1}(H_{k+1}\bx_{k+1}+G_{k+1}\bv_{k+1}+F_{k+1}\hat{\bx}_{k})^\T}\\\non
&\;\;-\EE{x_{k+1}((\EE{H_{k+1}}\EE{A_k}+\EE{F_{k+1}})\hat{x}_k)^\T}\\\label{Eq:derivation2:Gamma:xw:2}
&\;\;-\EE{x_{k+1}(\EE{H_{k+1}}\EE{B_k}u_k)^\T}.
\end{align}
Using the independence of $x_{k+1}$ and $v_{k+1}$, and canceling out identical terms,~\eqref{Eq:derivation2:Gamma:xw:2} becomes
\begin{align}\non
\Gamma_{x_{k+1}\tilde{y}_{k+1}}
&=\E[{\bx_{k+1}\bx_{k+1}^\T}]\E[{H_{k+1}^\T}]
    -\E[{\bx_{k+1}\hat{\bx}_{k}^\T}]\E[{A_k^\T}]\E[{H_{k+1}^\T}]\\\label{Eq:derivation2:Gamma:xw:3}
&\;\;-\E[{\bx_{k+1}}]u_k^\T\E[{B_k^\T}]\E[{H_{k+1}^\T}].
\end{align}
Before proceeding, we define $\Sigma_k\triangleq\E[{\bx_k\bx_k^\T}]$,
$\Delta_k\triangleq u_ku_k^\T$ and, in addition,
\begin{align}\label{Eq:derivation2:matrices:L}
&\Lambda_k\triangleq\E[{\hat{\bx}_k\hat{\bx}_k^\T}]=\E[{\hat{\bx}_k{\bx}_k^\T}]\\\label{Eq:derivation2:matrices:U}
&\Upsilon_k\triangleq\E[{x_k}]u_k^\T=\E[{\hat{x}_k}]u_k^\T,
\end{align}
where the RHS of~\eqref{Eq:derivation2:matrices:L} and~\eqref{Eq:derivation2:matrices:U} follow from the
orthogonality principle
and from the unbiasedness of $\hat{x}_k$, respectively. Note that $\Sigma_k$, $\Lambda_k$, and $\Delta_k$ are symmetric. \\
Using the independence of $\hat{x}_k$ and $w_k$,
\begin{align}\non
\EE{\bx_{k+1}\hat{\bx}_k^\T}
    &=\EE{(A_k\bx_k+B_ku_k+C_k\bw_k)\hat{\bx}_k^\T}\\\label{Eq:derivation2:x:x_hat}
    &=\EE{A_k}\Lambda_k+\EE{B_k}\Upsilon_k^\T,
\end{align}
which yields for~\eqref{Eq:derivation2:Gamma:xw:3}
\begin{align}\non
\Gamma_{x_{k+1}\tilde{y}_{k+1}}
&=\big(\Si_{k+1}
-(\E[{A_k}]\Lambda_k+\E[{B_k}]\Upsilon_k^\T)\E[{A_k^\T}]\\\label{Eq:derivation2:Gamma:xw:4}
&\;\;-\E[{\bx_{k+1}}]u_k^\T\E[{B_k^\T}]\big)\E[{H_{k+1}^\T}].
\end{align}
From~\eqref{Eq:problem:input:dynamics} we have
\begin{align}\non
\EE{\bx_{k+1}}
    &=\EE{A_k\bx_k+B_ku_k+C_k\bw_k}\\\label{Eq:derivation2:Ex}
    &=\EE{A_k}\EE{x_k}+\EE{B_k}u_k,
\end{align}
which, when substituted in~\eqref{Eq:derivation2:Gamma:xw:4}, leads to
\begin{align}\non
\Gamma_{x_{k+1}\tilde{y}_{k+1}}
&=\big(\Si_{k+1}
-(\E[{A_k}](\Lambda_k\E[{A_k^\T}]+\Upsilon_k\E[{B_k^\T}])\\\label{Eq:derivation2:Gamma:xw:5}
&\;\;+\E[{B_k}](\Upsilon_k^\T\E[{A_k^\T}]+\Delta_k\E[{B_k^\T}]))\big)\E[{H_{k+1}^\T}].
\end{align}
\subsection{Computation of $\Gamma_{\tilde{y}_{k+1}\tilde{y}_{k+1}}$}
\noindent
Since $\hat{y}_{k+1}^{-}$ is the LMMSE estimate of ${y}_{k+1}$ using $\mathcal{Y}_k$, $\tilde{y}_{k+1}$ is orthogonal to $\hat{y}_{k+1}^{-}$ and, using~\eqref{Eq:derivation2:time:update:y},
\begin{align}\non
&\Gamma_{\tilde{y}_{k+1}\tilde{y}_{k+1}}\\\non
    &\quad=\E[{(y_{k+1}-\hat{y}_{k+1}^-)y_{k+1}^\T}]\\\non
    &\quad=\E[{y_{k+1}y_{k+1}^\T}]-\E[{\hat{y}_{k+1}^-y_{k+1}^\T}]\\\non
    &\quad=\E[{y_{k+1}y_{k+1}^\T}]-(\E[{H_{k+1}}]\E[{A_k}]+\E[{F_{k+1}}])\E[{\hat{x}_ky_{k+1}^\T}]\\\label{Eq:derivation2:Gamma:ww:1}
    &\quad\;\;-\E[{H_{k+1}}]\E[{B_k}]u_k\E[{y_{k+1}^\T}].
\end{align}
Using \eqref{Eq:problem:input:measurement} and the independence of $\set{\hat{\bx}_k,\bx_{k+1}}$, $\set{H_{k+1},G_{k+1},F_{k+1}}$ and $v_{k+1}$, we have
\begin{align}\non
\E[{\hat{\bx}_{k}\by_{k+1}^\T}]
    &=\EE{\hat{\bx}_{k}(H_{k+1}\bx_{k+1}+F_{k+1}\hat{\bx}_{k})^\T}\\\label{Eq:derivation2:x_hat:y}
    &=\E[{\hat{\bx}_{k}\bx_{k+1}^\T}]\E[{H_{k+1}^\T}]
    +\Lambda_{k}\E[{F_{k+1}^\T}],
\end{align}
which, using~\eqref{Eq:derivation2:x:x_hat}, becomes
\begin{align}\non
\E[{\hat{\bx}_{k}\by_{k+1}^\T}]
    &=\Lambda_k(\E[{A_k^\T}]\E[{H_{k+1}^\T}]+\E[{F_{k+1}^\T}])\\\label{Eq:derivation2:x_hat:y:final}
    &\;\;+\Upsilon_k\E[{B_k^\T}]\E[{H_{k+1}^\T}].
\end{align}
\noindent
Due to the independence of $\set{\bx_{k+1},\hat{x}_k}$, $\bv_{k+1}$, and $\set{H_{k+1},G_{k+1}}$
\begin{align}\non
\E[{\by_{k+1}\by_{k+1}^\T}]
    &=\E[{H_{k+1}\bx_{k+1}\bx_{k+1}^\T H_{k+1}^\T}]+\E[{G_{k+1}\bv_{k+1}\bv_{k+1}^\T G_{k+1}^\T}]\\\non
    &\;\;+\E[{F_{k+1}\hat{\bx}_k{\hat{\bx}_k^\T F_{k+1}^\T}}]
    +\E[{H_{k+1}{\bx_{k+1}\hat{\bx}_k^\T}F_{k+1}^\T}]\\\label{Eq:derivation2:yy0}
    &\;\;+\E[{F_{k+1}{\hat{\bx}_k\bx_{k+1}^\T}H_{k+1}^\T}].
\end{align}
\nopagebreak[4]
Consider  
the last summand. From the smoothing property of the conditional expectation,
\begin{align}\non
\E[{F_{k+1}{\hat{\bx}_k\bx_{k+1}^\T}H_{k+1}^\T}]
    &=\EE{\E[{F_{k+1}{\hat{\bx}_k\bx_{k+1}^\T}H_{k+1}^\T}\mid{F_{k+1},H_{k+1}}]}\\\label{Eq:derivation2:FXH}
    &=\EE{F_{k+1}\E[{{\hat{\bx}_k\bx_{k+1}^\T}}]H_{k+1}^\T},
\end{align}
where we utilized the independence of $\set{H_{k+1},F_{k+1}}$ and $\set{\bx_{k+1},\hat{x}_k}$.\\
Similarly, since $\EE{x_{k+1}x_{k+1}^\T}=\Si_{k+1}$, $\EE{\bv_{k+1}\bv_{k+1}^\T}=I$, and $\EE{\hat{\bx}_k\hat{\bx}_k^\T}=\Lambda_k$, we obtain:
\begin{align}\label{Eq:derivation2:HSH}
&\E[{H_{k+1}\bx_{k+1}\bx_{k+1}^\T H_{k+1}^\T}]=\E[{H_{k+1}\Si_{k+1}H_{k+1}^\T}]\\\label{Eq:derivation2:GG}
&\E[{G_{k+1}\bv_{k+1}\bv_{k+1}^\T G_{k+1}^\T}]
    =\E[{G_{k+1}G_{k+1}^\T}]\\\label{Eq:derivation2:FLF}
&\E[{F_{k+1}\hat{\bx}_k{\hat{\bx}_k^\T F_{k+1}^\T}}]=\E[{F_{k+1}\Lambda_kF_{k+1}^\T}].
\end{align}
For future reference, we also note that
\begin{align}\label{Eq:derivation2:ASA}
&\E[{A_{k}\bx_{k}\bx_{k}^\T A_{k}^\T}]=\E[{A_{k}\Si_kA_{k}^\T}]\\\label{Eq:derivation2:ALB}
&\E[{A_{k}\bx_ku_k^\T B_{k}^\T}]=\E[{A_{k}\Up_kB_{k}^\T}]\\\label{Eq:derivation2:CC}
&\E[{B_{k}u_ku_k^\T B_{k}^\T}]=\E[{B_{k}\Delta_kB_{k}^\T}]\\\label{Eq:derivation2:CC}
&\E[{C_{k}\bw_k\bw_k^\T C_{k}^\T}]=\E[{C_{k}C_{k}^\T}].
\end{align}
Substituting~\eqref{Eq:derivation2:x:x_hat} in~\eqref{Eq:derivation2:FXH}, and using~\eqref{Eq:derivation2:FXH}-\eqref{Eq:derivation2:FLF} in~\eqref{Eq:derivation2:yy0},
\begin{align}\non
\E[{\by_{k+1}\by_{k+1}^\T}]
    &=\E[{H_{k+1}\Si_{k+1}H_{k+1}^\T}]+\E[{G_{k+1}G_{k+1}^\T}]\\\non
    &\;\;+\E[{F_{k+1}\La_{k}F_{k+1}^\T}]\\\non
    &\;\;+\EE{H_{k+1}(\EE{A_k}\Lambda_k+\EE{B_k}\Upsilon_k^\T)F_{k+1}^\T}\\\label{Eq:derivation2:yy}
    &\;\;+\EE{F_{k+1}(\Lambda_k\E[{A_k^\T}]+\Upsilon_k\E[{B_k^\T}])H_{k+1}^\T}.
\end{align}
In addition, we obtain, in a straightforward manner,
\begin{align}\non
\EE{y_{k+1}}
    &=(\EE{H_{k+1}}\EE{A_k}+\EE{F_{k+1}})\EE{x_k}\\\label{Eq:derivation2:Ey}
    &\;\;+\EE{H_{k+1}}\EE{B_k}u_k.
\end{align}
Using~\eqref{Eq:derivation2:HSH},~\eqref{Eq:derivation2:GG},
and~\eqref{Eq:derivation2:FLF} in~\eqref{Eq:derivation2:yy}, and
substituting~\eqref{Eq:derivation2:x_hat:y:final},~\eqref{Eq:derivation2:yy},
and~\eqref{Eq:derivation2:Ey} in~\eqref{Eq:derivation2:Gamma:ww:1} we
finally obtain
\begin{align}\non
&\Gamma_{\tilde{y}_{k+1}\tilde{y}_{k+1}}
=\EE{H_{k+1}\Si_{k+1}H_{k+1}^\T}+\EE{G_{k+1}G_{k+1}^\T}\\\non
    &\quad+\EE{F_{k+1}\La_kF_{k+1}^\T}-\EE{F_{k+1}}\La_k\EE{F_{k+1}}^\T\\\non
    &\quad-\EE{H_{k+1}}\EE{A_k}\La_k\EE{A_{k}^\T}\EE{H_{k+1}^\T}\\\non
    &\quad+\EE{H_{k+1}(\EE{A_k}\La_k+\EE{B_k}\Up_k^\T)F_{k+1}^\T}\\\non
    &\quad+\EE{F_{k+1}(\La_k\EE{A_k^\T}+\Up_k\EE{B_k^\T})H_{k+1}^\T}\\\non
    &\quad-\EE{H_{k+1}}\EE{A_k}\La_k\EE{F_{k+1}^\T}\\\non
    &\quad-\EE{F_{k+1}}\La_k\EE{A_k^\T}\EE{H_{k+1}^\T}\\\non
    &\quad-\EE{H_{k+1}}\EE{A_k}\Up_k\EE{B_{k}^\T}\EE{H_{k+1}^\T}\\\non
    &\quad-\EE{F_{k+1}}\Up_k\EE{B_{k}^\T}\EE{H_{k+1}^\T}\\\label{Eq:derivation2:Gamma:ww:3}
    &\quad-\EE{H_{k+1}}\EE{B_k}u_k\EE{y_{k+1}^\T}.
\end{align}
Notice, that a sufficient condition for the nonsingularity of $\Gamma_{\tilde{y}_{k+1}\tilde{y}_{k+1}}$ is $\E[{G_{k+1}G_{k+1}^\T}]\succ0$. To see this, recall that, by definition, $\Gamma_{\tilde{y}_{k+1}\tilde{y}_{k+1}}$ is positive semi-definite for any choice of $\EE{G_{k+1}G_{k+1}^T}$ and, in particular, for $G_{k+1}=0$.  But this means that the matrix on the RHS of~\eqref{Eq:derivation2:Gamma:ww:3} without $\EE{G_kG_k^T}$ is positive semi-definite, rendering $\EE{G_kG_k^T}\succ0$ a sufficient condition for the non-singularity of $\Gamma_{\tilde{y}_{k+1}\tilde{y}_{k+1}}$.

\subsection{Computation of the Second-Order Moments}
\noindent
Utilizing the independence of $x_k$, $w_k$ and $\set{A_k,B_k,C_k}$, and~\eqref{Eq:derivation2:ASA}--\eqref{Eq:derivation2:CC}, $\Si_{k+1}$ is given by
\begin{align}\non
\Si_{k+1}
&=\E[{\bx_{k+1}\bx_{k+1}^\T}]\\\non
  &=\EE{(A_k\bx_k+B_ku_k+C_k\bw_k)(A_k\bx_k+B_ku_k+C_k\bw_k)^\T} \\\non
     & = \E[{A_k\Si_kA_k^\T}]+\E[{A_k\Up_kB_k^\T}]+\E[{B_k\Up_k^\T A_k^\T}]\\\label{Eq:derivation2:Sigma}
     &\;\;+\E[{B_k\Delta_k B_k^\T}]+\E[{C_kC_k^\T}],
\end{align}
Next consider $\La_{k+1}$. 
Direct computation yields:
\begin{align}\non
\La_{k+1}
    &=\EE{\hat{\bx}_{k+1}{\bx}_{k+1}^\T}\\\non
    &=(L_k+K_k\EE{F_{k+1}})\E[{\hat{\bx}_{k}\bx_{k+1}^\T}]\\
    &\;\;+K_k\EE{H_{k+1}}\Si_{k+1}+J_ku_k\E[{{\bx}_{k+1}^\T}].
\end{align}
Using~\eqref{Eq:derivation2:x:x_hat} the latter becomes
\begin{align}\non
\La_{k+1}
&=
  (L_k+K_k\EE{F_{k+1}})(\La_k\E[{A_k^\T}]+\Up_k\E[{B_k^\T}])\\\label{Eq:derivation2:Lambda}
  &\;\;+J_k(\Up_k^\T\E[{A_k^\T}]+\Delta_k\E[{B_k^\T}])+K_k\E[{H_{k+1}}]\Si_{k+1}.
\end{align}
Finally,
$\Up_{k+1}=\EE{x_{k+1}}u_{k+1}^\T$.
Note that $\Delta_k$ is known for all $k$.
\subsection{Algorithm Summary}
\noindent
\begin{itemize}
\item[a)]{Initialization:}
$\hat{\bx}_0=\bar{\bx}_0$, $\Si_0=P_0+\bar{\bx}_0\bar{\bx}_0^\T$, $\La_0=\bar{\bx}_0\bar{\bx}_0^\T$, $\Up_0=\bar{\bx}_0u_0^\T$, $\Delta_0=u_0u_0^\T$.
\item[b)]{Recursion:}
For $k=1,2,\ldots$ perform the routine of Alg.~\ref{Alg:filter}.
\end{itemize}
\renewcommand{\algorithmicrequire}{\textbf{Input:}}
\renewcommand{\algorithmicensure}{\textbf{Output:}}
\begin{algorithm}[tbh]\caption{}
\begin{algorithmic}[1]
\REQUIRE{$\by_{k+1}$, $u_{k+1}$, $\hat{\bx}_k$, $\EE{x_k}$, $\Si_k$, $\La_k$, $\Up_k$, $\Delta_k$}%
\STATE{
Compute $\EE{A_k}$, $\EE{B_k}$, $\EE{C_{k}C_{k}^\T}$, $\EE{A_k\Si_kA_k^\T}$, $\EE{A_k\Up_kB_k^\T}$, and $\EE{B_k\Delta_kB_k^\T}$.}\label{alg:step:1}
\STATE{Compute $\EE{x_{k+1}}$ and $\Si_{k+1}$ using Eqs.~\eqref{Eq:derivation2:Ex} and~\eqref{Eq:derivation2:Sigma}.}\label{alg:step:2} %
\STATE{
Compute $\EE{H_{k+1}}$, $\EE{G_{k+1}G_{k+1}^\T}$, $\EE{F_{k+1}}$, $\EE{F_{k+1}\La_{k}F_{k+1}^\T}$, $\EE{H_{k+1}\Si_{k+1}H_{k+1}^\T}$, and $\EE{H_{k+1}(\EE{A_k}\La_k+\EE{B_k}\Up_k^\T)F_{k+1}^\T}$.}\label{alg:step:3}
\STATE{Compute $\Gamma_{x_{k+1}\tilde{y}_{k+1}}$ and $\Gamma_{\tilde{y}_{k+1}\tilde{y}_{k+1}}$ using Eqs.~\eqref{Eq:derivation2:Gamma:xw:5} and~\eqref{Eq:derivation2:Gamma:ww:3}.}\label{alg:step:4} %
\STATE{Compute $K_k$, $L_k$, and $J_k$ using Eqs.~\eqref{Eq:derivation2:recursive:Kk},~\eqref{Eq:derivation2:recursive:Lk}, and~\eqref{Eq:derivation2:recursive:Jk}, and $\hat{\bx}_{k+1}$ using Eq.~\eqref{Eq:problem:recursive}.}\label{alg:step:5} %
\STATE{Compute $\La_{k+1}$ using Eq.~\eqref{Eq:derivation2:Lambda} and $\Up_{k+1}$  by plugging $\EE{x_{k+1}}$ into~\eqref{Eq:derivation2:matrices:U}
.}\label{alg:step:6} %
\ENSURE{$\hat{\bx}_{k+1}$, $\EE{x_{k+1}}$, $\Si_{k+1}$, $\La_{k+1}$, $\Up_{k+1}$}%
\end{algorithmic}
\label{Alg:filter}
\end{algorithm}
Since the distribution of $\mathcal{M}_k$ is known, the expectations of steps~\ref{alg:step:1} and~\ref{alg:step:3} of Alg.~\ref{Alg:filter} may be calculated by, e.g., direct summations in case of discrete modes. In some cases, as demonstrated in Section~\ref{section:example1}, closed form expressions exist for the above expectations.

We note that the standard KF for a system with no inputs should be obtained when $\set{\mathcal{M}_k}$ is a deterministic sequence with $B_k=0$, $F_k=0$. In this setting we have 
\[
\Gamma_{x_{k+1}\tilde{y}_{k+1}}
    =(\Sigma_{k+1}-A_k\La_kA_k^\T)H_{k+1}^\T
 \]
and 
\[
\Gamma_{\tilde{y}_{k+1}\tilde{y}_{k+1}}
    =H_{k+1}(\Sigma_{k+1}-A_k\La_kA_k^\T)H_{k+1}^\T+G_{k+1}G_{k+1}^\T.
\]
 Substituting these in~\eqref{Eq:derivation2:recursive:thru:lemma} we indeed obtain the standard KF in the form where the time and measurement updates are combined together. The error covariances follow in a similar manner.

\subsection{Random Inputs}
\label{section:random:inputs}
In the second variant of~\eqref{Eq:problem:input:dynamics}, in which
$u_k=\hat{x}_k$, it turns out that the roles played by $A_k$ and $B_k$
are identical. Specifically, after replacing $u_k$ with
$\hat{\bx}_{k}$, at each step of the derivation of
Section~\ref{section:derivation}, $A_k$ and $B_k$ are multiplied by
the same quantities. Thus, the filter for the modified problem is
obtained from the one described in Alg.~\ref{Alg:filter} by replacing
$A_k$ with $A_k+B_k$ and nullifying $u_k$ and $\Upsilon_k$.
An alternative derivation, based on the orthogonality principle, may be found in~\cite{sigalov:modes:cdc2011}.


\section{Application to Target Tracking in Clutter}\label{section:example1}
In this section we demonstrate the proposed concept by casting the classical problem of tracking in clutter within our formulation, and applying the LMMSE filter of Section~\ref{section:derivation}.
%
%

\subsection{System and Clutter Models}
Consider a single target obeying a linear model. 
Setting  $A_k=A$, $B_k=0$, and $C_k={C}$ in~\eqref{Eq:problem:input:dynamics}
\begin{align}\label{Eq:application1:dynamics}
\bx_{k+1}&=A\bx_k+C\bw_k.
\end{align}
Here $A$ and $C$ are deterministic matrices, accounting for the state dynamics and process noise covariance, respectively, and $\set{w_k}$ is a scalar process noise sequence. The target state is observed via the the equation
\begin{align}\label{Eq:application1:true_measurement}
{\by}_{k}^{\rm true}&=\Hnom\bx_k+\Gnom{\bv}_{k}^{\rm true},
\end{align}
where ${\bv}_{k}^{\rm true}$ represents measurement noise. In addition, at each time, a number of clutter detections are obtained. These will be denoted as $\{{\by}_{k,i}^{\rm cl}\}_{i=1}^{N-1}$, where $N$ is the total number of detections. Clutter measurements 
do not carry any information about the target of interest. They are, however, indistinguishable from true detections in the sense that they carry information of the same type (say, position). At each time, the clutter measurements are assumed to be independent of each other, of the clutter measurements at other times, and of the true state and observation. In addition, we assume that they are uniformly distributed in space.
To correctly model the distribution of the clutter detections, we note
that, typically, at each scan, the sensor initiates a validation window
centered about the predicted target position, and the algorithm processes
only those measurements obtained within the window.
Since the clutter detections are uniformly distributed in space, they
are also uniformly distributed within the validation window.

We define the measurement vector $\by_k$ to be the
concatenation of all measurements from time $k$, $N-1$ of which correspond to
clutter, and one originating from the true target. The location of the
true measurement within this concatenated vector is, of course,
unknown to the algorithm.
This setting can be modeled using~\eqref{Eq:problem:input:measurement} by letting the mode $\cM_k$ be distributed as
\begin{align}\non
\cM_k &= \set{H_k,G_k,F_k}\\\label{Eq:application1:mode}
&=
\begin{cases}
\set{
\paren{\begin{smallmatrix}\Hnom\\ {0}\\\vdots\\ {0}\end{smallmatrix}},
\diag\paren{\begin{smallmatrix}\Gnom\\ \Gcl  \\\vdots\\ \Gcl  \end{smallmatrix}},
\paren{\begin{smallmatrix}{0}\\ \Hnom A\\\vdots\\ \Hnom A\end{smallmatrix}}
}, & \text{w.p. } \frac{1}{N}\\
\hspace{0.9cm}\vdots\hspace{2cm}\vdots\hspace{1.6cm}\vdots\\
\set{
\paren{\begin{smallmatrix}{0}\\\vdots\\ {0}\\\Hnom\end{smallmatrix}},
\diag\paren{\begin{smallmatrix}\Gcl  \\\vdots\\ \Gcl  \\ \Gnom  \end{smallmatrix}},
\paren{\begin{smallmatrix}\Hnom A\\\vdots\\ \Hnom A\\{0}\end{smallmatrix}}
}, & \text{w.p. } \frac{1}{N},
\end{cases}
\end{align}
where $\Gcl$ is the square-root of the covariance matrix associated with the clutter.

For example, the first realization of $\set{H_k,G_k,F_k}$
in~\eqref{Eq:application1:mode} corresponds to the
scenario in which the first of the $N$ observations is the true target
measurement, ${\by}_{k}^{\rm true}$, generated according
to~\eqref{Eq:application1:true_measurement}, while the other $N-1$
measurements are clutter, each of which is generated according to
\begin{align}\label{Eq:application1:clutter_measurement}
{\by}_{k,i}^{\rm cl}&=\Hnom A\hat{\bx}_{k-1}+\Gcl{\bv}_{k,i}^{\rm cl},\quad i=2,\ldots,N.
\end{align}
Here, $\Hnom A\hat{\bx}_{k-1}$ is the predicted true measurement at
time $k$, which is also the center of the validation window
, so that clutter measurements at time $k$ are
uniformly distributed around this quantity. Namely, ${\bv}_{k,i}^{\rm cl}$ has a uniform distribution.
The overall number of measurements in the validation window, $N$, is assumed to be known,
but may vary in time. Thus, the dimensions of $H_k$, $G_k$, and $F_k$
may depend on $k$.

It is readily observed that the matrices $\set{H_k,G_k,F_k}$ are correlated in this setting. This renders the approach of~\cite{dekoning:stochastic:1984} inapplicable in the current scenario. Furthermore, it can be seen that without the feedback term in the measurement equation, it is impossible to account for the fact that clutter is uniformly distributed in a window centered about the predicted measurement. In fact, any linear method disregarding this term, such as \cite{costa1994lmm,dekoning:stochastic:1984}, must assume that clutter measurements are distributed about $0$.

Notice that we assumed, for simplicity, that the true measurement is
always present in the validation window.  To account for the possibility that the true measurement does not fall in the validation window, the option
\[
\set{H_k,G_k,F_k}=\set{
\bm{0},I_N\otimes\Gcl,\bm{1}_N\otimes\Hnom A
}
\] 
needs to be added to the set of possible realizations
in~\eqref{Eq:application1:mode}. Here, $\otimes$ stands for the Kronecker product, $\bm{1}_N$ is
an $N\times1$ vector comprising all ones, and $I_N$ is the $N\times N$
identity matrix. %
The probability of this outcome is $(1-P_D)(1-P_G)$ where $P_D$ is the probabilty of target detection, assumed known, and $P_G$ is the probability that, upon target detection, the true measurement falls in the validation window. This parameter is defined by the user and, typically, it affects the window size as discussed in the sequel.
Note that, when no measurements are available, $N=0$,
and~\eqref{Eq:problem:recursive} becomes (at the absence of $u_k$) $\hat{\bx}_{k+1}=L_k\hat{\bx}_{k}$, which
corresponds to a simple prediction (time update) without consecutive measurement update, as expected.
\subsection{Matrix Computations}
To invoke the algorithm presented in Section~\ref{section:derivation}
we need to compute the expectations of  Steps~\ref{alg:step:1} and~\ref{alg:step:3} of
Alg.~\ref{Alg:filter}. Although these may be evaluated numerically,
via direct summations, in the present example closed-form expressions
exist, as we show next for the simple setting in which the true measurement is always present in the validation window (extensions are straightforward.)
\indent As the matrices of the dynamics equation are deterministic,
$\EE{A_k}=A$, $\EE{B_k}={0}$, $\EE{C_{k}C_{k}^\T}=CC^\T$,
$\EE{A_k\Up_kB_k^\T}={0}$, $\EE{B_k\Delta_kB_k^\T}={0}$, and
$\EE{A_k\Si_kA_k^\T}=A\Si_kA^\T$.
Also, according to the distribution defined in~\eqref{Eq:application1:mode},
\begin{align}\label{Eq:application1:matrices2}
 &\EE{H_{k+1}}=\frac{1}{N}\bm{1}_N\otimes\Hnom\\
 &\EE{F_{k+1}}=\frac{N-1}{N}\bm{1}_N\otimes\Hnom A.
\end{align}
The remaining terms read
\begin{align}\label{Eq:application1:matrices3}
\E[{H_{k+1}\Si_{k+1}H_{k+1}^\T}]
    &=\frac{1}{N}I_N\otimes\Hnom \Si_{k+1}\Hnom^\T\\\label{Eq:application1:matrices4}
\E[{G_{k+1}G_{k+1}^\T}]
    &=\frac{1}{N}I_N\otimes\paren{\Gnom\Gnom^\T+(N-1)\Gcl\Gcl^\T}\\\label{Eq:application1:matrices5}
\EE{F_{k+1}\La_{k}F_{k+1}^\T}&=\Xi\otimes\paren{\Hnom A\La_kA^\T\Hnom^\T},
\end{align}
where
\begin{align}\label{Eq:application1:matrices6}
\Xi=
\begin{cases}
\frac{1}{N}\paren{(N-2)\bm{1}_N\bm{1}_N^\T+I_N}, & N>1\\\
0, &N=1.
\end{cases}
\end{align}
Finally,
\begin{align}\non
&\EE{H_{k+1}(\EE{A_k}\La_k+\EE{B_k}\Up_k^\T)F_{k+1}^\T}\\\label{Eq:application1:matrices7}
&=\frac{1}{N}(\bm{1}_N\bm{1}_N^\T-I_N)\otimes\paren{\Hnom A\La_kA^\T\Hnom^\T}.
\end{align}

The spatial distribution of clutter is uniform in the validation window, whose size determines $\Gcl\Gcl^\T$.

\subsection{Discussion}
It is easy to see that, in the present case, $\Gamma_{\tilde{y}_{k+1}\tilde{y}_{k+1}}=I_N\otimes D$ where
\begin{align}\non
D
&=\frac{1}{N}\Hnom A\La_kA^\T\Hnom^\T+\frac{1}{N}\Hnom\Si_{k+1}\Hnom^\T\\\non
&\;\;+\frac{1}{N}\Gnom\Gnom^\T
    +\frac{N-1}{N}\Gcl\Gcl^\T.
\end{align}
Moreover,
\begin{align}\non
\Gamma_{x_{k+1}\tilde{y}_{k+1}}
    &=(\Si_{k+1}-A\La_kA^\T)\EE{H_{k+1}^\T}\\
    &=\frac{1}{N}(\Si_{k+1}-A\La_kA^\T)\paren{\Hnom^\T\cdots\Hnom^\T}^\T
,
\end{align}
and
\begin{align}\non
K_k &= \Gamma_{x_{k+1}\tilde{y}_{k+1}}\Gamma_{\tilde{y}_{k+1}\tilde{y}_{k+1}}^{-1}\\
&=\frac{1}{N}\bm{1}_N^\T\otimes\paren{(\Si_{k+1}-A\La_kA^\T)\Hnom^\T D^{-1}}.
\end{align}
Since $y_{k+1}$ is a concatenation of all the
observations from time $k+1$, the product $K_k\by_{k+1}$
in~\eqref{Eq:problem:recursive} is the average of these
measurements, pre-multiplied by
$(\Si_{k+1}-A\La_kA^\T)\Hnom^\T D^{-1}$. Consequently, the LMMSE estimator
for tracking a target in clutter is a KF-like algorithm, operating on the
average of all detections in the validation window. In this respect,
its mode of operation resembles classical methods. For example,
the probabilistic data association (PDA)~\cite{pda:barshalom:1975} method implements a KF driven by the weighted average of all measurements in the window,
and the nearest neighbor (NN) filter~\cite{MMT} is a KF driven by the
measurement nearest to the prediction assigning it a weight of $1$ and assigning $0$ to the rest of the measurements.
%
%

\subsection{Numerical Study}\label{section:numerical}
We consider a one-dimensional tracking scenario, in which the state comprises position and velocity information, $\bx_k=(p_k\quad v_k)^\T$. Starting at $\bx_0\sim\gaus{\bar{x}_0}{P_0}$ with $\bar{x}_0=(0\quad0)^\T$ and $P_0=30I_2$, the target is simulated for $400$ time
units using~\eqref{Eq:application1:dynamics} with $A=\paren{\begin{smallmatrix}1 & 0.2\\0 & 0.95\end{smallmatrix}}$ and $C=\frac{1}{2}\paren{\begin{smallmatrix}1/2 \\1\end{smallmatrix}}$. The process and measurement noises are taken to be Gaussian.
The true measurement is generated using~\eqref{Eq:application1:true_measurement} with $\Hnom=(1\quad 0)$ and $\Gnom=\sqrt{30}$.
%
The target is detected with probability $P_D=0.95$ and the probability that the true observation falls in the validation window is taken to be $P_G=0.99$. A validation window is set about the predicted measurement position. Its size, $d$, is determined to comply with $P_G$ (see~\cite[p.130]{MMT} for details). Once the window is determined, the clutter variance of~\eqref{Eq:application1:clutter_measurement} is $\Gcl\Gcl^\T=d^2/12$.

The derived algorithm is compared with NN and PDA filters, that are equipped with the same windowing logic and parameters. All algorithms are initialized with $\hat{x}_0=\bar{x}_0$ and the initial error covariance matrix is taken to be $P_0$.
When dealing with tracking in clutter, using the MSE
as the only performance measure may result in misleading
conclusions, since, eventually the estimate will draw away from the true
measurement and follow the clutter, and the errors will become meaninglessly large.
We thus use two measures of performance to evaluate the algorithms.
The first is the time until the target is lost, defined as
the third consecutive time when the measurement of a detected target falls outside the validation window. The second measure is the root MSE (RMSE) calculated over the time interval until the first of the three algorithms loses track.

\begin{figure*}
\begin{center}
  \subfloat{\includegraphics[width=0.4\textwidth, trim=0.5cm
    0cm 1cm 0.5cm, clip]{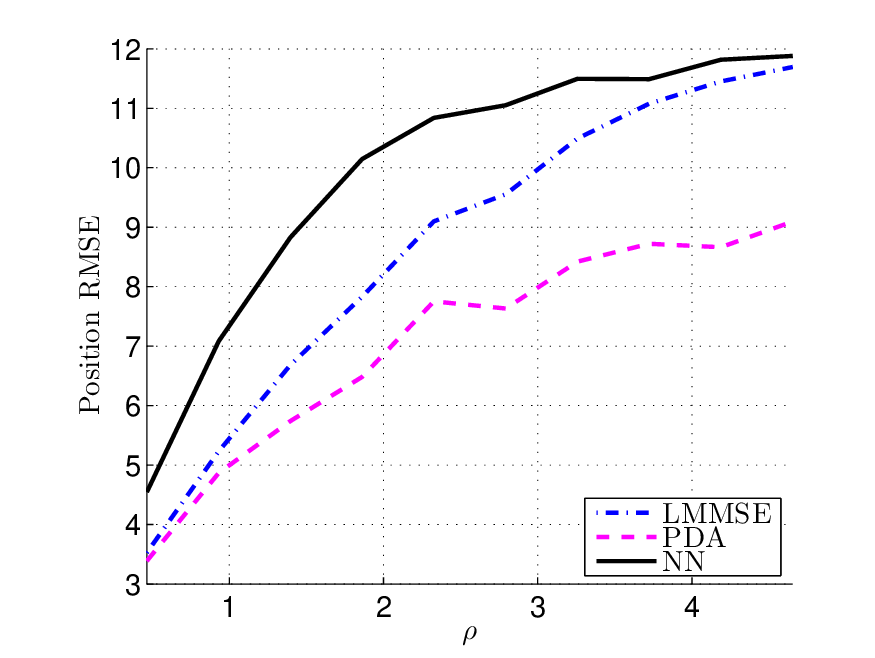}\label{fig:clutter:position}}\qquad
  \subfloat{\includegraphics[width=0.4\textwidth, trim=1cm 0cm
    1cm 0.5cm,
    clip]{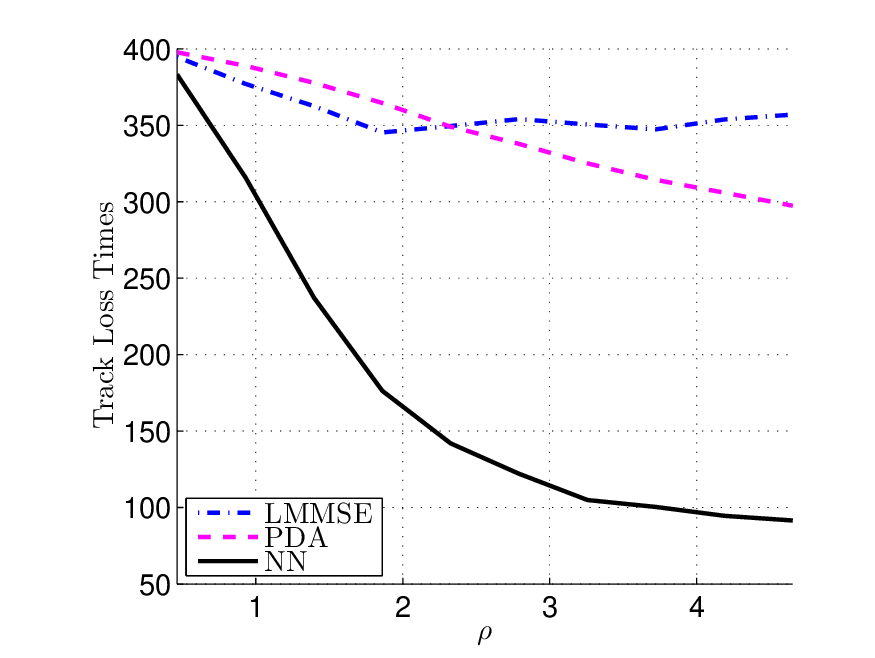}\label{fig:clutter:times}}
  \caption{Position RMSE (left) and track loss time (right) vs. clutter density.}
\label{fig:clutter}
\end{center}
\end{figure*}

We test the algorithms at a range of clutter densities. Let $\rho$ to be the average number of clutter measurements
falling in an interval of one standard deviation of the (true)
measurement noise. Averaged over $1000$ independent Monte Carlo runs,
the average position RMSE and track loss times are plotted, versus
$\rho$, in Fig.~\ref{fig:clutter}.
%
%
It is readily seen that the LMMSE filter attains competitive performance relatively to the nonlinear algorithms. Specifically, for heavy clutter regimes it maintains longest track loss times. It is not very surprising that the errors of PDA are better, since these are calculated before the first of the three algorithms has lost track (NN in all cases). During this period the PDA performs a more efficient, nonlinear manipulation on the measurements. However, for high clutter rates, it is probable that clutter measurements will be assigned higher weights than the true detection, eventually leading to a track loss. In this case, it is better to simply average the measurements, as the linear filter does.




\section{Conclusion}\label{section:conclusion}
We proposed a new formulation of JLS, where the dynamics and
measurement equations are allowed to depend on previous estimates of
the state representing closed-loop control input and measurement validation window.
We derived an LMMSE recursive algorithm for this setting, and
illustrated the approach in the context of tracking in
clutter. In this case, our filter demonstrates competitive
performance, when compared with classical, nonlinear methods.
\appendices


\bibliographystyle{IEEEtran}
\bibliography{../../../Common_Files/IEEEabrv,../../../Common_Files/SigalovReferences}

\end{document}